\documentstyle[psfig,times]{mn}

\def\etal{et~al.}
\def\spose#1{\hbox to 0pt{#1\hss}}
\def\lta{\mathrel{\spose{\lower 3pt\hbox{$\mathchar"218$}}
     \raise 2.0pt\hbox{$\mathchar"13C$}}}
\def\gta{\mathrel{\spose{\lower 3pt\hbox{$\mathchar"218$}}
     \raise 2.0pt\hbox{$\mathchar"13E$}}}
\def\Ha{H$\alpha$}

\def\iraf{{\sc iraf}}
\def\noao{{\sc noao}}

\def\kms{\,km\,s\,$^{-1}$}
\def\Ho50{$H_0 = 50$km\,s$^{-1}$\,Mpc$^{-1}$}

\title[Redshifts for radio sources from BRL]{The final two redshifts for
radio sources from the equatorial BRL sample}

\author[P.~N.~Best \etal]{P.~N.~Best,$^1$ J.~A.~Peacock,$^1$ 
M.~H.~Brookes,$^1$ R.~E.~Dowsett,$^1$, H.~J.~A.~R{\"o}ttgering,$^2$\\
\\
\LARGE J.~S.~Dunlop,$^1$ M. D. Lehnert$^3$\\
$^1$ Institute for Astronomy, Royal Observatory Edinburgh,
Blackford Hill, Edinburgh EH9 3HJ, UK\\
$^2$ Sterrewacht Leiden, Postbus 9513, 2300 RA Leiden, the Netherlands\\
$^3$ Max-Planck-Institut f{\"u}r extraterrestrische Physik, Postfach 1312, 
85741 Garching, Germany}

\pagerange{\pageref{firstpage}--\pageref{lastpage}}
\pubyear{1999}
\begin{document}
\label{firstpage}

\maketitle

\begin{abstract}
\noindent 
Best, R{\"o}ttgering and Lehnert (1999, 2000a) defined a new sample of
powerful radio sources from the Molonglo Reference Catalogue, for which
redshifts were compiled or measured for 177 of the 178 objects. For the final
object, MRC1059-010 (3C249), the host galaxy is here identified using
near-infrared imaging, and the redshift is determined from VLT
spectroscopy. For one other object in the sample, MRC0320+053 (4C05.14), the
literature redshift has been questioned: new spectroscopic observations of
this object are presented, deriving a corrected redshift. With these two
results, the spectroscopic completeness of this sample is now 100\%.

New redshifts are also presented for PKS0742+10 from the Wall \& Peacock
2.7\,GHz catalogue, and PKS1336+003 from the Parkes Selected
Regions. PKS0742+10 shows a strong neutral hydrogen absorption feature in its
Lyman-$\alpha$ emission profile.
\end{abstract}


\section{Introduction}
The role of radio sources in astrophysical and cosmological studies is wide
and varied (e.g. see McCarthy 1993 for a review)\nocite{mcc93}, and the
utility of spectroscopically complete samples of radio sources cannot be
understated. By far the best studied sample of radio sources is the revised
3CR sample \cite{lai83}, which is a spectroscopically complete sample of the
brightest 173 radio sources in the northern sky, selected at 178\,MHz ($S_{\rm
178 MHz} > 10.9$\,Jy). The inaccessibility of this sample to large southern
hemisphere telescope facilities prompted Best \etal\ (1999; hereafter
BRL99)\nocite{bes99e} to define a roughly equivalent sample of equatorial radio
sources from the Molonglo Reference Catalogue (MRC; Large \etal\
1981)\nocite{lar81}, according to the criteria $S_{\rm 408 MHz} \ge 5$\,Jy,
$-30^{\circ} \le \delta \le +10^{\circ}$, $|b| \ge 10^{\circ}$. The reader is
referred to BRL99 for a complete description of the sample and its properties.

This equatorial radio source sample, hereafter referred to as BRL, consists of
178 objects. BRL99 compiled or measured redshifts for 174 of these, and Best
\etal\ \shortcite{bes00a} determined a further three, leaving just the radio
source MRC1059$-$010 without measured redshift. Subsequently the redshift of
MRC0320+053, obtained by BRL99 from the NASA/IPAC Extragalactic Database (NED),
has also been called into question \cite{dev98}. In this paper, a new host
galaxy identification is obtained for MRC1059$-$010, based on near--infrared
imaging observations, and spectroscopic redshifts are provided for both
MRC1059$-$010 and MRC0320+053. In addition, redshifts are presented for two
further radio sources from samples with high spectroscopic completeness.

\section{Observations and Data Reduction}
\label{obs}

\subsection{Infrared imaging observations}

MRC1059$-$010 was observed on the night of 22nd April 2002 with the United
Kingdom InfraRed Telescope (UKIRT) using the UKIRT Fast Track Imager (UFTI).
UFTI is a 1024 by 1024 pixel array with a plate scale of 0.091 arcsec per
pixel. A total of 18 minutes of data were taken in the K--band, corresponding
to two cycles of a 9--point dither pattern with 15 arcsecond offsets, with a
60 second exposure at each position. The data were reduced using standard
techniques within the \iraf\ \noao\ reduction software (see e.g. Best \etal\
2003 for details)\nocite{bes03a}. The combined K--band image is shown with the
contours of the radio emission overlaid in Figure~\ref{1059kfig}. Regular
observations of standard stars throughout the night showed that conditions
were photometric, with a K--band zero-point magnitude of $22.33 \pm 0.02$.

\subsection{VLT spectroscopic observations}

Spectroscopic observations of MRC1059$-$010 were obtained during morning
twilight of the night of 28th February 2003, using the FORS1 instrument of
ESO's Very Large Telescope (VLT). The observations were made using the 300V+10
grating, which provides a spatial scale of 0.2 arcsec, a spectral resolution
of 12\AA\ through a 1-arcsec slit, and a useful wavelength coverage of about
3600 to 8000\AA. Two 1200 second spectra were obtained; see Table~\ref{alltab}
for details.

During this same observing run, observations of two further radio sources
without spectroscopic redshifts were also carried out at times when the
primary targets were not accessible (see Table~\ref{alltab}). These sources
are PKS0742+10, one of the few remaining sources without redshifts in the
2.7\,GHz all sky radio source sample of Wall \& Peacock \shortcite{wal85}, and
PKS1336+003, drawn from the Parkes Selected Region sample \cite{dun89b}.

The data were reduced using standard packages within \iraf. The different
exposures were combined and one-dimensional spectra were extracted from an
angular extent of 2.0 arcsec. These were then wavelength calibrated using
observations of HeNeAr arc lamps. After correction for the airmass, flux
calibration was achieved using observations of the spectrophotometric standard
stars Hiltner 600, GD108 and Feige 56.

\subsection{NTT spectroscopic observations}

The radio source MRC0320+053 (RA: 03 20 41.54, Dec: 05 23 34.6; B1950) was
observed using EMMI on the New Technology Telescope (NTT), during evening
twilight on the night of 3 March 2003. The observations were made through a
1.5-arcsecond slit using Grism number 4, which provided a useful wavelength
range of about 6000 to 9500\AA\ at a spectral resolution of $\sim 15$\AA. The
CCD was binned 2 by 2 on readout, giving a spatial scale of 0.33 arcsec per
binned pixel. Three five-minute exposures were obtained, but the first two of
these had a significantly higher background level.

Data reduction followed the same procedure as for the VLT observations except
that, because of the varying background level, to obtain maximum
signal--to--noise the three individual frames were given weights proportional
to the inverse of their background counts in the image combination procedure.
The one--dimensional spectrum was extracted from an angular extent of 3.3
arcsec along the slit and the star LTT3218 was used for flux calibration.

\begin{table*}
\caption{\label{alltab} Details of the spectroscopic observations. Note that
no equivalent widths are provided for the emission lines of MRC1059-010 due to
the failure to detect any significant continuum emission. Note also that the
line flux quoted for the Ly$\alpha$ emission of PKS0742+10 is that measured in
the spectrum, and has not been corrected for the HI absorption.}
\begin{tabular}{llcrcrrrrr}
\hline
\multicolumn{1}{c}{Source}     & 
\multicolumn{1}{c}{Obs.}       & 
\multicolumn{1}{c}{Telescope}  & 
\multicolumn{1}{c}{Exp.}       &  
\multicolumn{1}{c}{Redshift}   & 
\multicolumn{1}{c}{Emis.}      & 
\multicolumn{1}{c}{$\lambda$}  &
\multicolumn{1}{c}{Line  Flux} &
\multicolumn{1}{c}{Eq. width}  &
\multicolumn{1}{c}{Vel. width}   \\  

                               &  
\multicolumn{1}{c}{Date}       & 
\multicolumn{1}{c}{\& Instrum.}& 
\multicolumn{1}{c}{time}       &
                               &
\multicolumn{1}{c}{line}       &
\multicolumn{1}{c}{(obs.)}     &
\multicolumn{1}{c}{$\times 10^{19}$} &
\multicolumn{1}{c}{(rest-frame)}      &
\multicolumn{1}{c}{(deconv)}   \\

                                 &
\multicolumn{1}{c}{[d/m/y]} &
                                 &
\multicolumn{1}{c}{[s]}          &
                                 &
                                 &
\multicolumn{1}{c}{[\AA]}        &
\multicolumn{1}{c}{[W\,m$^{-2}$]} &
\multicolumn{1}{c}{[\AA]}        &
\multicolumn{1}{c}{[km\,s$^{-1}$]} \\
\hline
MRC1059-010 & 28/02/03 & VLT/FORS1 & $2 \times 1200$ & $1.554 \pm 0.004$  &
        CIV  1549    & 3962.9 & $ 0.13 \pm 0.04$ & ---\hspace*{0.5cm} & $ 500 \pm 500$ \\
&&&&&  CIII] 1909    & 4874.4 & $ 0.22 \pm 0.02$ & ---\hspace*{0.5cm} & $1500 \pm 300$ \\
&&&&&   CII] 2326    & 5948.4 & $ 0.16 \pm 0.03$ & ---\hspace*{0.5cm} & $ 750 \pm 400$ \\
&&&&& [NeIV] 2425    & 6178.8 & $ 0.22 \pm 0.07$ & ---\hspace*{0.5cm} & $ 600 \pm 500$ \\
MRC0320+053 & 03/03/03 & NTT/EMMI  & $3 \times 300$ & $0.1785 \pm 0.0002$ &
      [NII] 6548     & 7717.4 & $ 171 \pm 20~~~$ &  33\hspace*{0.5cm} & $1250 \pm 150$  \\ 
&&&&& H$\alpha$ 6563 & 7735.5 & $  30 \pm ~~5~~~$&   6\hspace*{0.5cm} & $1250 \pm 150$  \\
&&&&& [NII] 6583     & 7758.6 & $ 470 \pm 20~~~$ &  92\hspace*{0.5cm} & $1250 \pm 150$  \\
&&&&& [SII] 6717     & 7915.1 & $  41 \pm 10~~~$ &   8\hspace*{0.5cm} & $ 900 \pm 300$  \\
&&&&& [SII] 6731     & 7931.7 & $  28 \pm ~~8~~~$&   5\hspace*{0.5cm} & $ 900 \pm 300$  \\
PKS0742+10  & 27/02/03 & VLT/FORS1 & $2 \times 1200$ & $2.624 \pm 0.003 $    &
     Ly$\alpha$ 1216 & 4409.7 & $2.86 \pm 0.29$  & 162\hspace*{0.5cm} & $3450 \pm 350$ \\
&&&&&   CIV  1549    & 5607.9 & $0.05 \pm 0.02$  &   2\hspace*{0.5cm} & $ 700 \pm 500$ \\
&&&&&   HeII 1640    & 5947.9 & $0.31 \pm 0.05$  &  17\hspace*{0.5cm} & $1150 \pm 550$ \\
&&&&&  CIII] 1909    & 6919.5 & $0.42 \pm 0.05$  &  21\hspace*{0.5cm} & $ 950 \pm 300$ \\
PKS1336+003 & 27/02/03 & VLT/FORS1 & $2 \times 900$ & $1.236 \pm 0.002 $    &
        HeII 1640    & 3663.6 & $0.35 \pm 0.10$  &  12\hspace*{0.5cm} & $ 600 \pm 500$ \\
&&&&&  CIII] 1909    & 4265.7 & $0.53 \pm 0.06$  &  42\hspace*{0.5cm} & $ 500 \pm 300$ \\
&&&&&   CII] 2326    & 5207.6 & $0.35 \pm 0.04$  &  35\hspace*{0.5cm} & $1450 \pm 500$ \\
&&&&& [NeIV] 2425    & 5427.9 & $0.16 \pm 0.03$  &  14\hspace*{0.5cm} & $ 600 \pm 400$ \\
&&&&& [NeV]  3426    & 7669.5 & $0.15 \pm 0.04$  &   9\hspace*{0.5cm} & $ 711 \pm 500$ \\
\hline

\end{tabular}
\end{table*}

\begin{figure}
\centerline{
\psfig{file=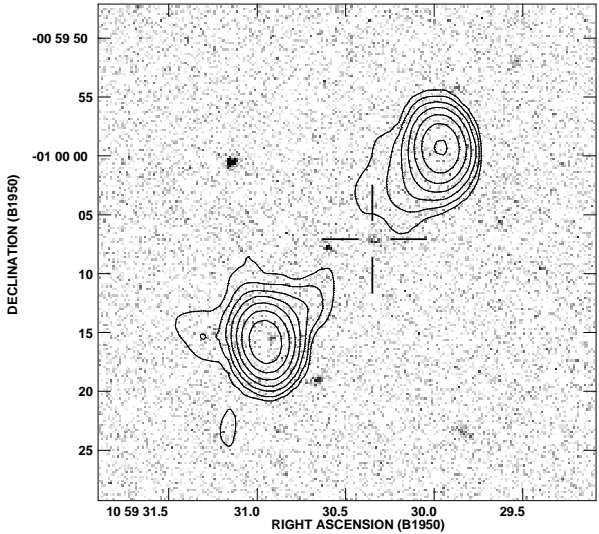,angle=-90,width=8.2cm,clip=}
}
\caption{\label{1059kfig} The K-band image of MRC1059-010, with contours of
radio emission overlaid. Contours are at 1.5 $\times$ (1,2,4,8,16,32,64,128)
mJy/beam. There are two potential host galaxies identified. The true host (as
confirmed by spectroscopy) is marked by the crosshairs, has a position of 10
59 30.33, $-$01 00 07.2 (B1950) and a magnitude of $K = 18.9 \pm 0.3$
(3-arcsec diameter aperture).}
\end{figure}

\begin{figure}
\centerline{
\psfig{file=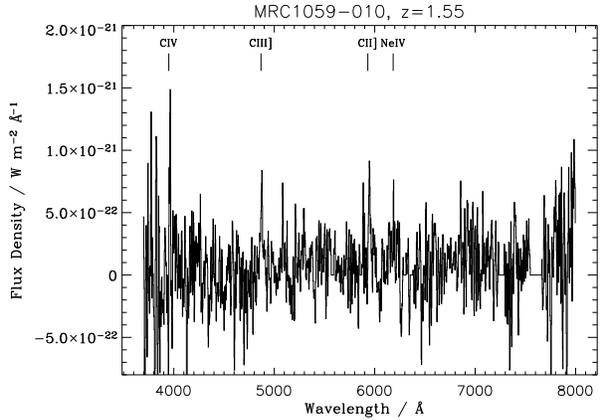,angle=90,width=8.2cm,clip=}
}
\caption{\label{1059specfig} The VLT/FORS1 spectrum of MRC1059-010. Note 
that although some of the emission lines may not be convincing on this
one-dimensional spectrum, all are clearly real when viewed on the
two-dimensional data.}
\end{figure}

\section{Results}

\subsection{The BRL sample}

Two possible host galaxies for MRC1059$-$010 are clearly identified on the new
K--band image (Figure~\ref{1059kfig}), the more westerly of these being more
likely host galaxy based upon its location. The position (RA: 10 59 30.33,
Dec: $-$01 00 07.2; B1950) of this galaxy is offset by over an arcsecond from
that determined by BRL99 based upon a tentative 4$\sigma$ detection of an $R
\sim 24$ galaxy. No K--band counterpart to that $R$--band object is seen,
suggesting that it was simply a noise peak. The VLT spectroscopic slit was
aligned to include both of the potential K--band host galaxies, and 4 strong
emission lines were detected towards the favoured western candidate (see
Figure~\ref{1059specfig}), identifying this galaxy as the radio source host
and placing the radio source at redshift $z=1.55$. Details of the emission
line properties are provided in Table~\ref{alltab}; both the line luminosities
and the emission line ratios are typical of distant radio galaxies (cf Best
\etal\ 2000b)\nocite{bes00c}.  The K--band magnitude of this galaxy is $K=
18.9 \pm 0.3$ as measured through a 3-arcsec diameter aperture; this is nearly
a magnitude fainter than a typical powerful radio source at this redshift
\cite{bre02a}, which partially explains the difficulty in obtaining an optical
identification. The host galaxy also has $R-K \gta 5$, making it one of the
reddest powerful radio galaxies at those redshifts (cf Dunlop \etal\ 1989).
\nocite{dun89a}

\begin{figure}
\centerline{
\psfig{file=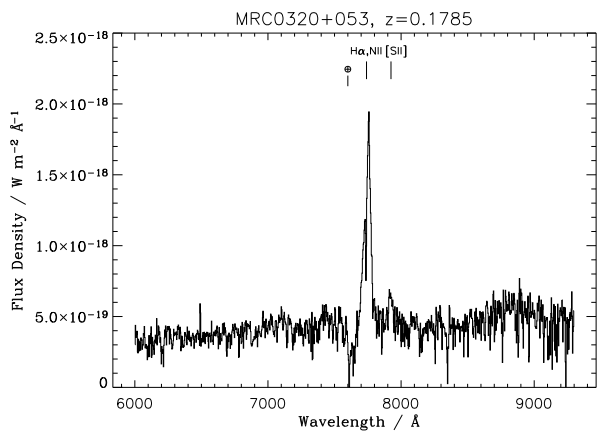,angle=90,width=8.2cm,clip=}
}
\caption{\label{0320specfig} The NTT/EMMI spectrum of MRC0320+053. The
absorption feature marked at 7600\AA\ is atmospheric.}
\end{figure}

The redshift of MRC0320+053 has also been securely determined through the
detection of strong $H\alpha$ and NII emission lines with confirming [SII]
(Figure~\ref{0320specfig}). The new redshift of $z=0.1785$ is considerably
lower than the value of $z=0.575$ which had erroneously appeared in NED, and
is more in line with that estimated by de Vries \etal\ \shortcite{dev98} from
the K-band magnitude. To derive the emission line properties the [NII] and
\Ha\ emission lines were deblended by fixing the relative wavelengths and
forcing the width of each line to be the same, but allowing the redshift of
the system and the relative line intensities to vary; despite the apparent
closeness, the 7600\AA\ atmospheric absorption feature does not significantly
influence this fit. The intensity ratio of the two [NII] lines ought to be 3;
this value was not fixed in the fitting procedure, but the best-fit is
consistent with this, giving confidence in the results. The [SII] doublet was
similarly deblended, and the measured line ratio is consistent with gas in the
low density limit, $n_e \lta 100$cm$^{-3}$ \cite{ost89}. This radio galaxy has
a [NII]/\Ha\ ratio in excess of 10, which corresponds to a remarkably high
ionisation, well above that of typical AGN \cite{ost89}. This may be related
to the fact that this is a compact radio source, smaller than 0.2 arcsec,
with the emission lines arising from very close to the active nucleus.

\subsection{Parkes Radio Sources}

\begin{figure*}
\begin{tabular}{cc}
\psfig{file=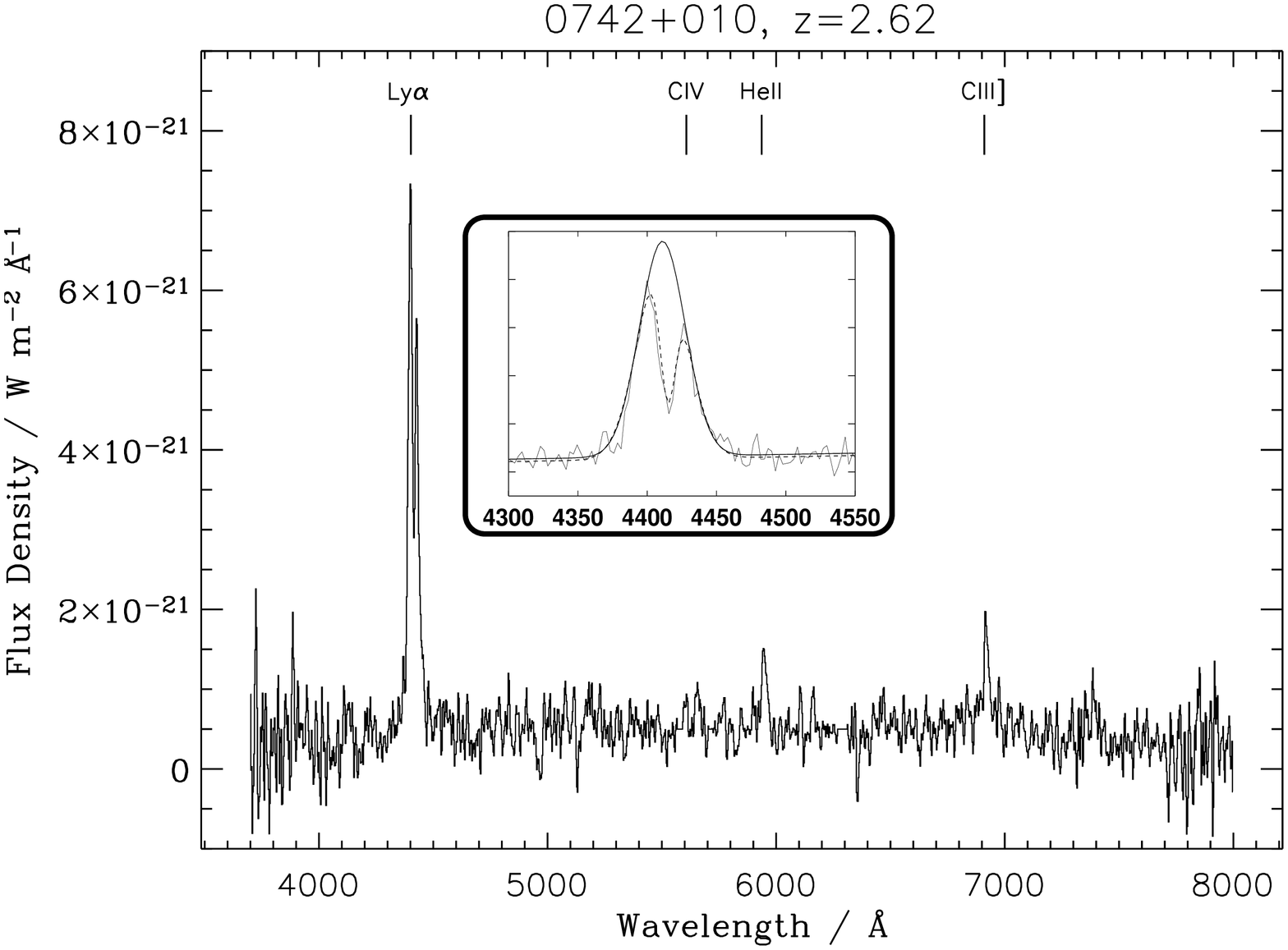,angle=-90,width=7.9cm,clip=}
&					   
\psfig{file=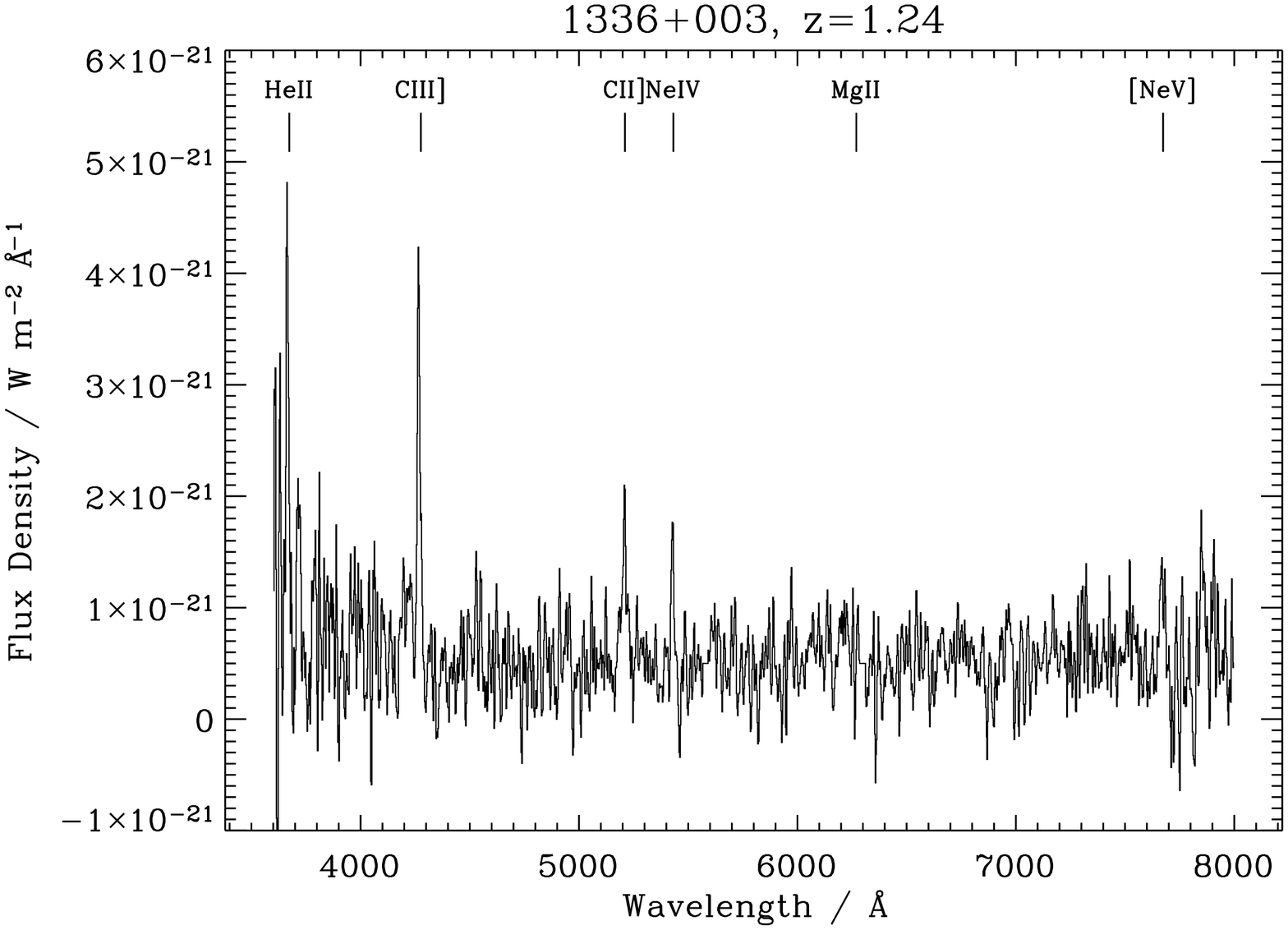,angle=90,width=8.2cm,clip=}
\\
\end{tabular}
\caption{\label{pksspecfigs} The VLT/FORS1 spectra of PKS0742+10 (left) and
PKS1336+003 (right). The inset in the PKS0742+10 spectrum shows an enlargement
of the region around the Ly-$\alpha$ emission line, showing that the data
(thin solid line) is well matched by a model (dashed line) constructed from a
Gaussian emission line (thicker solid line) together with a Voigt absorption
profile.}
\end{figure*}

The extracted spectra of the two Parkes radio sources are shown in
Figure~\ref{pksspecfigs}, and details of their emission line properties are
provided in Table~\ref{alltab}. Emission lines are detected for both objects,
confirming their optical identifications. PKS0742+10 shows a strong absorption
feature within its Ly-$\alpha$ emission; this is reminiscent of the results of
van Ojik \etal\ \shortcite{oji97}, who found neutral hydrogren absorption
features to be almost ubiquitous in small ($\lta 50$\,kpc) high redshift radio
sources. PKS0742+10 has a radio size of only a few milliarcsec
\cite{sta01}, fitting this trend.

The absorption feature was modelled using a Gaussian emission line together
with a Voigt absorption profile, each convolved with the spectral resolution
of the observations (see inset in Figure~\ref{pksspecfigs}). The absorption
profile is centred at 4412.6\AA, corresponding to a 150\kms\ redshift compared
to the mean velocity determined from the other emission lines. If it is
assumed that this absorption is optically thin then the neutral hydrogen
column density can be calculated from the equivalent width of the absorption,
giving $\rm{N(HI)} = (4 \pm 1) \times 10^{15}$cm$^{-2}$, with a velocity width
of $b=30 \pm 10$\,km\,s$^{-1}$. This column density will be a lower limit if
the absorber is optically thick: as discussed by Dopita \& Sutherland
\shortcite{dop03}, in the optically thick regime the column density cannot be
reliably estimated from low spectral resolution observations unless the
absorption is strong enough to be damped, which is not the case here. This has
been well demonstrated by the results of Jarvis \etal\ \shortcite{jar03}, who
took high spectral resolution observations of the radio galaxy 0200+015 and
found remarkably different absorption column densities to those found by van
Ojik \etal\ \shortcite{oji97} for the same source at much lower resolution
(two absorbers with $4 \times 10^{14}$cm$^{-2}$ column density instead of one
with $\sim 10^{19}$cm$^{-2}$).

\subsection{Concluding remarks}

By obtaining the final optical identification and the final two spectroscopic
identifications of sources from the BRL sample, a fully spectroscopically
complete catalogue of radio sources has been constructed at equatorial
declinations, complementing the northern 3CR sample. This is currently the
only complete sample of powerful radio sources which is accessible to southern
hemisphere telescopes, making it a unique and powerful resource in radio
source astrophysics.

Although the Parkes Selected Regions still only have about 50\% spectroscopic
completeness, a search through NED reveals that the Wall \& Peacock 2.7\,GHz
sample has now reached a spectroscopic completeness of 224 out of 233
sources. The nine remaining sources without redshifts are PKS0008--42,
PKS0316+16, PKS0407--65, PKS1308--22, PKS1600+33, PKS1740--51, PKS2008--06,
PKS2032--35 and PKS2150--52.
  
\section*{Acknowledgements} 

PNB would like to thank the Royal Society for financial support through its
University Research Fellowship scheme. MHB and RED are grateful for the
support of PPARC research studentships. The authors thank Omar Almaini for
carrying out the observations of MRC0320+053. The United Kingdom Infrared
Telescope is operated by the Joint Astronomy Centre on behalf of the
U.K. Particle Physics and Astronomy Research Council. The spectroscopic
observations were carried out using the ESO Very Large Telescope at the
Paranal observatory under Program-ID Number 071.A-0622(A), and the ESO New
Technology Telescope at the La Silla observatory under Program-ID Number
70.B-0747 .  This research has made use of the NASA/IPAC Extragalactic
Database (NED) which is operated by the Jet Propulsion Laboratory, California
Institute of Technology, under contract with NASA. The authors thank the 
referee for his/her efficiency in reviewing the manuscript.

\bibliography{pnb} 
\bibliographystyle{mn}

\label{lastpage}
\end{document}